\documentclass[iop]{emulateapj}
\usepackage{apjfonts}

\usepackage[tight]{subfigure}
\usepackage{amsmath}
\usepackage{graphicx}
\usepackage{wasysym}
\usepackage{color}
\usepackage{verbatim}

\newcommand{\nuc}[2]{\ensuremath{\mathrm{^{#1}#2}}}

\newcommand{\msun}{\ensuremath{\mathrm{M}_\odot}}

\def\dm15{$\Delta m_{15}(B)$}

\def\lesssim{\mathrel{\hbox{\rlap{\hbox{\lower4pt\hbox{$\sim$}}}\hbox{$<$}}}}

\def\gtrsim{\mathrel{\hbox{\rlap{\hbox{\lower4pt\hbox{$\sim$}}}\hbox{$>$}}}}

\shorttitle{A unified merger model for normal and rapidly declining
                  Type Ia Supernovae}
\shortauthors{R.~Pakmor et al.}

\begin{document}

\title{Helium-ignited violent mergers as a unified model for normal 
and rapidly declining Type~Ia Supernovae}

\author
{
 R.~Pakmor$^{1,3}$,
 M.~Kromer$^{2,3}$,
 S.~Taubenberger$^{2}$
 and V.~Springel$^{1,4}$
}

\altaffiltext{1}
{
  Heidelberger Institut f\"{u}r Theoretische Studien, 
  Schloss-Wolfsbrunnenweg 35, 
  D-69118 Heidelberg, Germany
}

\altaffiltext{2}
{
  Max-Planck-Institut f\"ur Astrophysik,
  Karl-Schwarzschild-Str. 1, 
  D-85748 Garching, Germany
} 

\altaffiltext{3}
{
  Kavli Institute for the Physics and Mathematics of the Universe,
  University of Tokyo, Kashiwa, Chiba 277-8583, Japan
}

\altaffiltext{4}
{
  Zentrum f\"ur Astronomie der Universit\"at Heidelberg,
  Astronomisches Recheninstitut, M\"{o}nchhofstr. 12-14, 69120
  Heidelberg, Germany
}

\begin{abstract}
  The progenitors of Type Ia Supernovae (SNe~Ia) are still unknown,
  despite significant progress during the last years in theory and
  observations. Violent mergers of two carbon--oxygen (CO) white
  dwarfs (WDs) are one candidate suggested to be responsible for at
  least a significant fraction of normal SNe~Ia. Here, we simulate the
  merger of two CO WDs using a moving-mesh code that allows for the
  inclusion of thin helium (He) shells (0.01\,\msun) on top of the WDs,
  at an unprecedented numerical resolution. The accretion of He onto the
  primary WD leads to the formation of a detonation in its He
  shell. This detonation propagates around the CO WD and sends a
  converging shock wave into its core, known to robustly trigger a second
  detonation, as in the well-known double-detonation scenario for
  He-accreting CO WDs. However, in contrast to that scenario where
  a massive He shell is required to form a detonation through thermal instability,
  here the He detonation is ignited dynamically.  Accordingly the
  required He-shell mass is significantly smaller, and hence its
  burning products are unlikely to affect the optical display of the
  explosion. We show that this scenario, which works for CO primary
  WDs with CO- as well as He-WD companions, has the potential to
  explain the different brightness distributions, delay times and
  relative rates of normal and fast declining SNe~Ia.  Finally, we
  discuss extensions to our unified merger model needed to obtain a
  comprehensive picture of the full observed diversity of SNe~Ia.
\end{abstract}


\section{Introduction}
\label{sec:intro}

Historically, Type Ia supernovae (SNe~Ia) have been believed to be a
uniform class of objects. However, over the past two decades large
supernova surveys have revealed that this class is in fact rather
inhomogeneous.  Nonetheless, until recently, it has been argued that
most -- if not all -- SNe~Ia originate from delayed detonations of
Chandrasekhar-mass carbon--oxygen (CO) White Dwarfs (WDs) accreting
hydrogen-rich material from a non-degenerate companion star
\citep[e.g.][]{mazzali2007a}.

Lately, however, an increasing number of observational constraints
have challenged this idea. In particular, the very nearby normal SN~Ia
2011fe did not show any signs of a companion star
\citep{li2011b,bloom2012a}, nor any signature of hydrogen in nebular
spectra \citep{shappee2012a} which is expected in the
single-degenerate scenario owing to material stripped from the
companion \citep{marietta2000a,leonard2007a}.
Moreover, there is a growing number of SN~Ia remnants in which no
remaining companion star could be found \citep[e.g.][]{schaefer2012a,
  hernandez2012a}.  Also, population-synthesis studies predict that
the expected rate of single-degenerate Chandrasekhar-mass explosions
is at least a factor of a few below the observed rate (e.g.\
\citealt{ruiter2009a, mennekens2010a}, but see \citealt{han2004a}).

Double-degenerate systems, in contrast, avoid most of these problems:
there is no surviving companion star expected and they can naturally
explain the absence of hydrogen features in late-time spectra.  At the
same time, recent first-principle three-dimensional simulations of
violent mergers of two CO WDs reproduce subluminous and normal SNe~Ia
well \citep{pakmor2010a,pakmor2012a,roepke2012a}.  However, it is
debated whether the temperature and density conditions on the surface 
of the WD, reached in the simulations, are sufficient to form a carbon 
detonation \citep{pakmor2010a,dan2011a,raskin2012a,pakmor2012a}.

Here, we present a new three-dimensional simulation of the merger of
two massive CO~WDs using, for the first time in supernova modelling, 
the moving-mesh code \textsc{arepo} \citep{springel2010a}. 
\textsc{arepo} allows for an highly adaptive resolution and thus
the inclusion of thin helium (He) shells on top of the CO WDs. We 
show how this presence of He in the binary system leads to a robust
explosion mechanism for violent mergers.

In Sect.~\ref{sec:merger} we
present our new simulation.  We then discuss the implications of the
new ignition mechanism that naturally leads to two distinct
populations of SNe~Ia in Sect.~\ref{sec:display}.
Sect.~\ref{sec:popsyn} presents the expected brightness distributions,
rates and delay times of both populations in the context of
observational constraints.  
We conclude with a brief summary and outlook in Sect.~\ref{sec:summary}.

\begin{figure*}
  \centering
  \includegraphics[width=\linewidth]{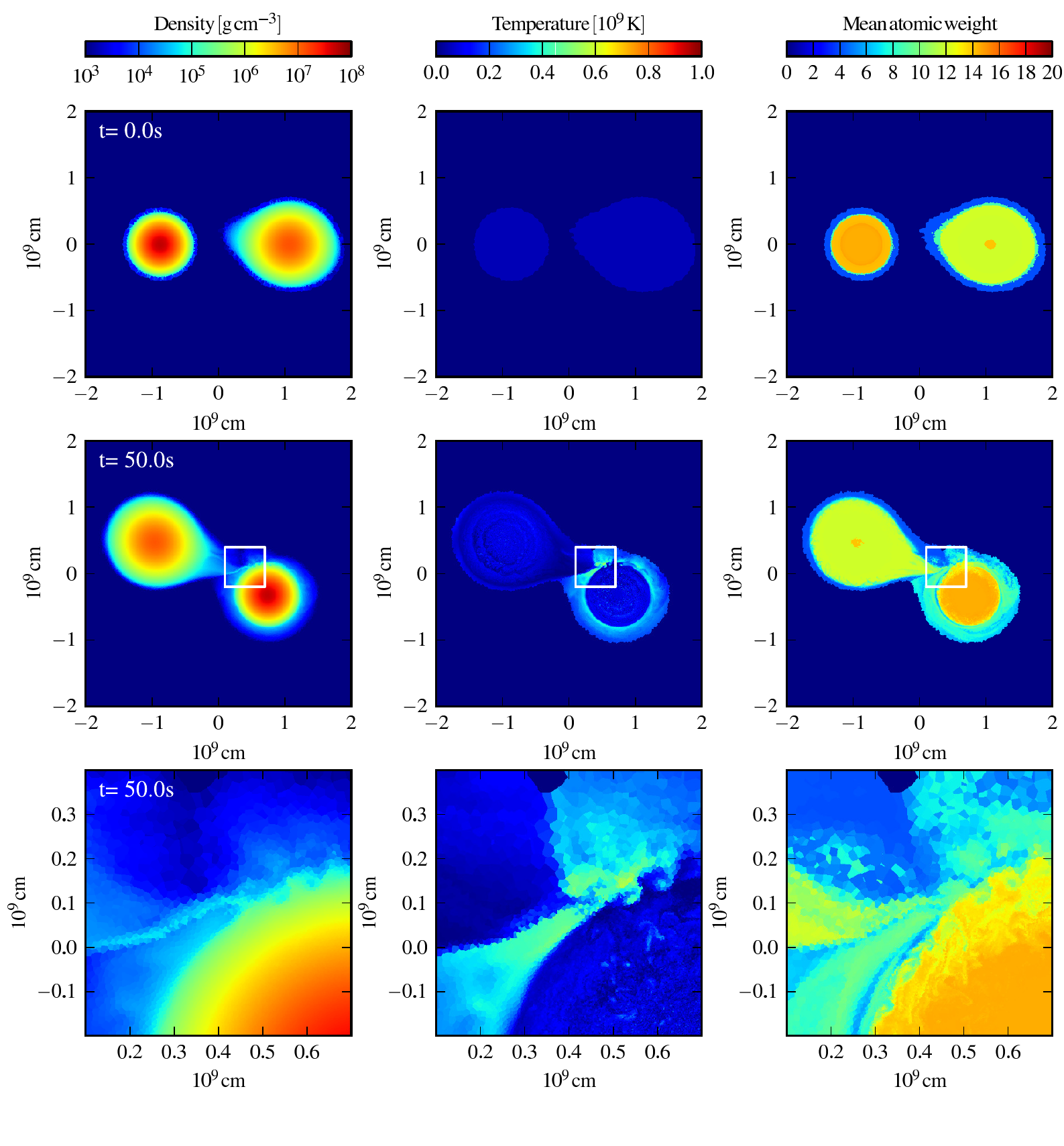}
  \caption{Time evolution of density, temperature and mean atomic
    weight (columns from left to right) slices in the orbital plane for our simulation 
    of the merger of a $1.1\,\msun$ with a $0.9\,\msun$ CO WD (shown are
    slices in the plane of rotation). The bottom row shows enlargements 
    of the area marked by the white boxes in the second row.}
  \label{fig:merger}
\end{figure*}

\begin{figure*}
  \centering
  \includegraphics[width=\linewidth]{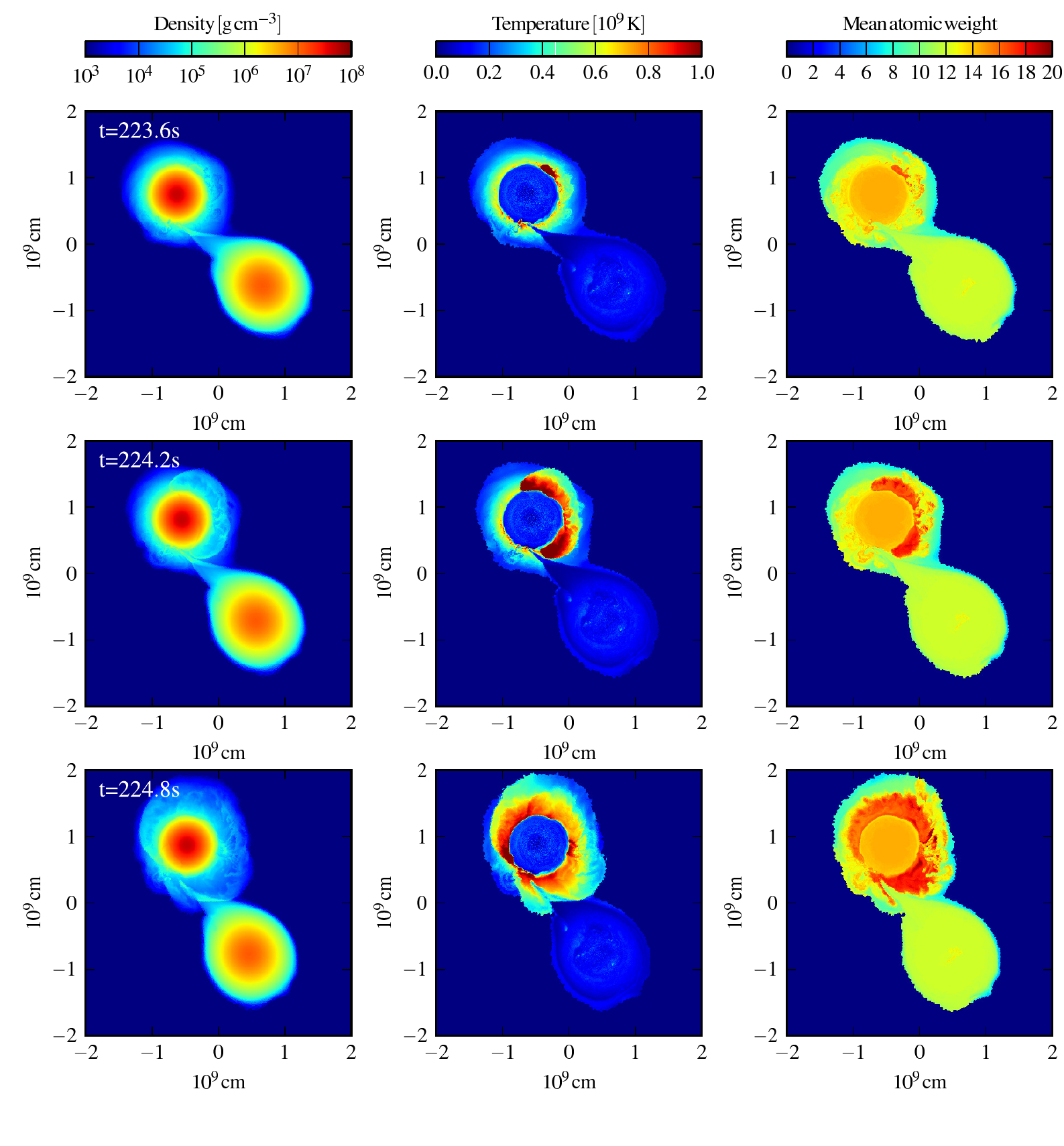}
  \caption{As Fig.~\ref{fig:merger}, but showing the later stages of 
    the merger, when the He detonation forms on the surface of the 
    primary.}
  \label{fig:merger2}
\end{figure*}

\begin{figure*}
  \centering
  \includegraphics[width=\linewidth]{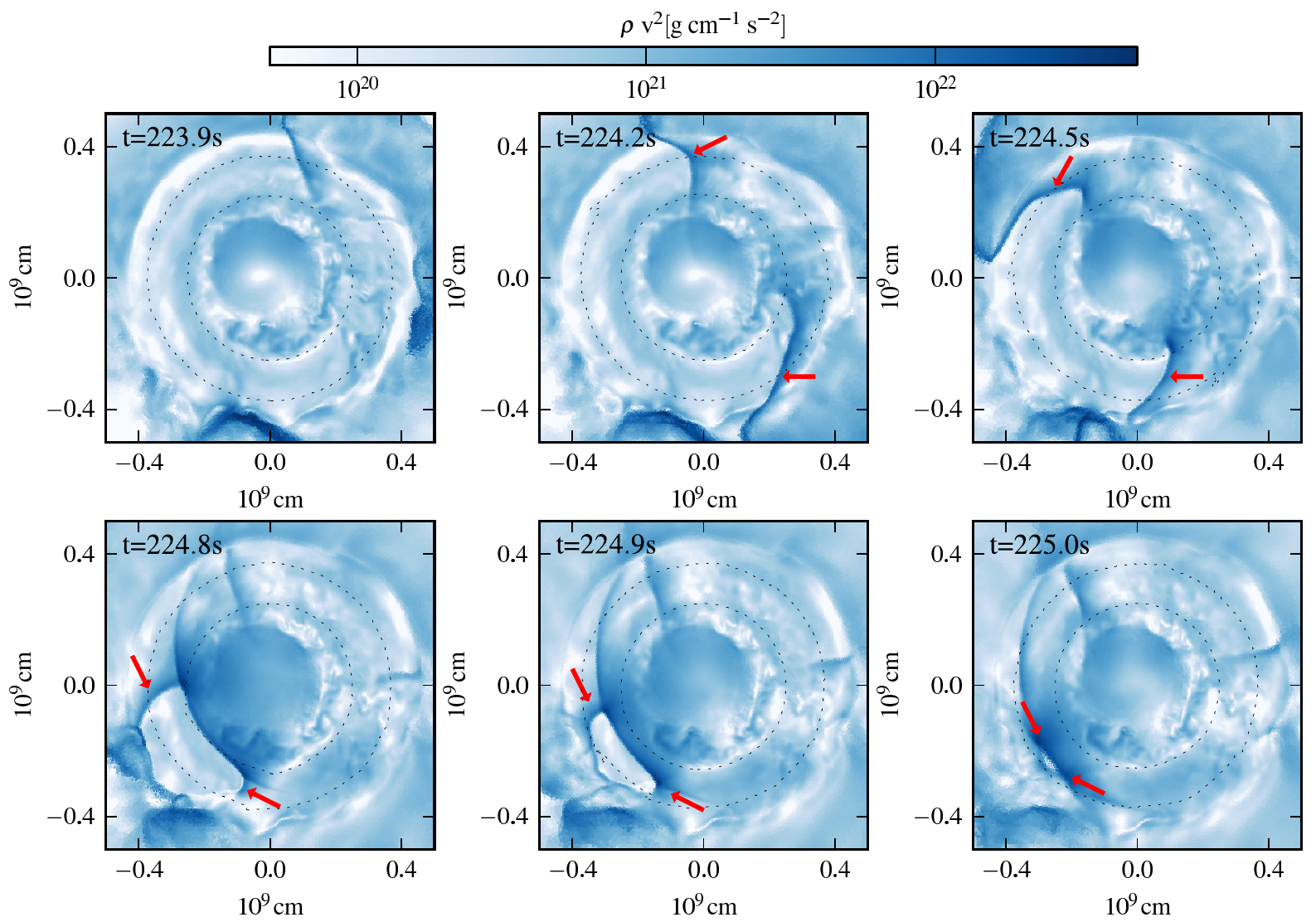}
  \caption{Slices of specific kinetic energy of the primary WD material 
    after the orbital velocity of the binary system and 
    an approximate solid-body spin of the primary WD have 
    been removed (dotted black lines show density contours for 
    $2\times10^6\,\rm{g\,cm^{-3}}$ and $10^7\,\rm{g\,cm^{-3}}$).
    A shock (marked by red arrows) emerges from the He detonation, 
    propagates around the primary and converges in its CO core.}
  \label{fig:shock}
\end{figure*}

\section{Helium-ignited violent mergers}
\label{sec:merger}
We simulate the merger of a $1.1\,\msun$ CO~WD and a $0.9\,\msun$
CO~WD using the moving-mesh code \textsc{arepo} \citep{springel2010a},
which we extended with the same degenerate equation of state and 
nuclear-reaction network as described in \citet{pakmor2012b}. 
Specifically, in our simulation we use a $13$-isotope $\alpha$-network
which is active for cells with temperature $>10^6\,
\mathrm{K}$, unless $\nabla \cdot \textbf{v} < 0$ and $\left| 
\nabla P \right| r_{\mathrm{cell}} / P > 0.33$ (following 
\citealt{seitenzahl2009b}).

We operate \textsc{arepo} in the moving-mesh mode \citep{springel2010a},
i.e.\ mesh-generating points move with the fluid velocity of their 
respective cell and a small correction to keep the grid regular. In 
addition, we employ explicit refinement to keep the mass of the cells 
within a factor of two of a given target mass and to guarantee a maximum 
volume difference between neighboring cells of a factor of $10$. 
In this mode, \textsc{arepo} is almost Lagrangian and significantly 
reduces advection errors compared to static grid codes. At the same 
time, \textsc{arepo} does not suffer from artificial viscosity, 
sampling noise and the large errors in gradient estimates typical of
SPH codes.

\subsection{Setup}
We construct binary initial conditions the same way as in 
\citet{pakmor2012a}, using the \textsc{gadget} code
\citep{springel2005a,pakmor2012b}. We map the binary initial
conditions to \textsc{arepo} by converting the SPH particles to 
Voronoi-cells that carry the mass and momentum of the former
SPH particles. We set the temperature to $5\times10^{7}\,\mathrm{K}$ 
for all cells to remove temperature noise introduced 
in SPH. The binary is placed at the center of a $10^{12}\,\mathrm{cm}$
wide box with low background density ($10^{-5}\,\mathrm{g\,cm^{-3}}$).

In contrast to \citet{pakmor2012a}, we use composition profiles
from \citet{iben1985a} to construct the two WDs. To
be able to resolve the He shell on the surface of both WDs, we
increase the amount of He for both WDs to a total of $0.01\,\msun$.
This is done by changing the composition in the outermost 
$0.005\,\msun$ of the WDs to pure He and the following $0.01\,\msun$
to a composition of  $50\%$ He, $30\%$ C and $20\%$ O.

We start the actual {\sc arepo} simulation with a mass resolution of  $1.1\times 
10^{-7}\,\msun$ everywhere. During the first second, we decrease the 
target mass of the cells in the He shell linearly with time by a factor 
of $10$ to a final mass resolution of $1.1\times 10^{-8}\,\msun$
at $t=1\,\mathrm{s}$. The He shell is tracked using a passive scalar.

\subsection{Simulation results}
\label{sec:merger_results}
The initial setup and evolution is shown in Fig.~\ref{fig:merger}. As 
mass and angular momentum are transferred to the primary WD, it heats up at
the surface and its outer layers start to spin relative to its core. This
leads to the development of Kelvin-Helmholtz instabilities between
the He shell and the CO core of the primary WD. Over time, the He
is mixed with CO material and after a few orbits there is no pure He
layer anymore.
Moreover, He starts to burn episodically on the surface of the
primary, but does not form a detonation at first. 

As the orbit of the
binary system shrinks, mass transfer becomes faster and more violent.
After about $7$ orbits, when the orbital period has decreased to about 
$30\,\mathrm{s}$, a He detonation forms close to the surface of the
CO core of the primary WD in a region compressed by a shock in the 
He layer. Before nuclear burning starts, this region has a temperature 
of $\sim5\times 10^8\,\mathrm{K}$, a density of $\sim2.5\times 10^5\,
\mathrm{g\,cm^{-3}}$ and a composition of roughly equal parts of He, 
C and O by mass. It is resolved by cells with a radius of 
$30$--$40\,\mathrm{km}$.

Note that the detonation forms dynamically, similar to previous results by 
\citet{guillochon2010a} for systems with high He accretion rates on 
CO primaries, rather than by thermal instabilities in 
stably-accreting steadily-burning systems \citep[see, e.g.,][]{shen2010a}.
This implies that a He detonation can form for smaller He shells than 
in steadily-He-burning systems \citep{bildsten2007a} as long as 
the accretion rate becomes high enough to ignite the He shell 
dynamically and there is sufficient He in the outer layers of the primary 
WD to form a detonation at all \citep[see, e.g.,][]{moll2013a}. This likely
includes most CO~WD primaries with outer He shells and binaries where 
the companion can transfer enough He to the primary before the system
merges, in particular for He~WD companions.

In our simulation, the He detonation propagates around the primary
WD's surface (Fig.~\ref{fig:merger2}), sending a shock wave into its 
CO core, that  converges in one point in the core (Fig.~\ref{fig:shock}). 
From two- and three-dimensional simulations sub-Chandrasekhar-mass
explosions \citep{fink2007a,fink2010a,moll2013a}, we know that such 
a converging shock inevitably leads to the formation of a subsequent 
C detonation. In our three-dimensional full merger simulation, however, 
we cannot resolve the convergence region well enough to follow the 
ignition of this detonation directly.

In our simulation the secondary WD is still intact at the time the detonation 
emerges, which makes it significantly different from previous simulations of 
similar systems that included a comparably massive He shell
\citep{raskin2012a}. In those simulations, the He detonation only ignites
when the secondary WD is already in the process of merging, and the He
detonation is localized and does not propagate around the primary WD.
The differences are most likely the result of different hydro schemes and
the significantly higher resolution we use in the He shell (factor 
$\sim 100$ in mass). However, it is possible that the time when He detonates 
depends on the masses of the two WDs and that systems with low-mass
primaries and smaller accretion rates only detonate during the 
actual merger.

\section{The observational display of violent mergers}
\label{sec:display}

The explosion mechanism described above directly suggests that there
are at least two different populations of SNe~Ia, which originate from
explosions in pairs of CO WDs and CO/He WDs, respectively.

In both populations, the amount of \nuc{56}{Ni} synthesized in the
explosion, and therefore the intrinsic brightness, correlates directly
with the mass of the CO primary WD, since the central
density of the secondary WD is always below the threshold for
\nuc{56}{Ni} production at the time of detonation. However, depending
on the nature of the secondary WD (He or CO), the total mass of the ejecta
changes. Moreover, the interaction of the ashes of the primary 
with the secondary WD may lead to its destruction, either by subsequent 
nuclear burning in the secondary or, for low-mass WDs, by the impact 
of the shock-wave alone. In both cases, the density profile and composition 
structure of the ejecta are changed, which directly affects observables.

Carbon-ignited violent mergers of two CO WDs have already
been shown to reproduce the light curves and spectra of normal SNe~Ia
fairly well \citep{pakmor2012a,roepke2012a}. Whether this holds for
He-ignited violent mergers has to be investigated by detailed
simulations. In He-accreting sub-Chandrasekhar-mass
double-detonation models the spectral properties are highly sensitive
to the burning products of the He shell
\citep[e.g.][]{kromer2010a}. However, in these models it is assumed
that the He shell grows massive enough to be ignited by a thermal
instability \citep{bildsten2007a}. In contrast, 
in our merger model, for a given primary WD mass the He shell is 
less massive. Moreover, the He shell may be significantly mixed with 
CO material (see Sect.~\ref{sec:merger_results}), preventing the He from 
synthesizing iron-group elements in the detonation (in our simulation the
helium detonation synthesizes only $2\times10^{-8}\,\msun$ of elements heavier
than Calcium). Therefore, it is conceivable that in our scenario the ashes
of the He shell will have little or no effect on spectra and light curves.

It has also been shown that mergers of two low-mass CO WDs
reproduce colors and spectra of objects similar to SN~1991bg very
well \citep{pakmor2010a}. However, such mergers fail to reproduce the
narrow, fast evolving light curves of 1991bg-like SNe. The light
curve width of SNe~Ia is mainly set by the amount of \nuc{56}{Ni}
synthesized during the explosion since the \nuc{56}{Ni} provides
a significant part of the total opacity of the ejecta. As a consequence
brighter explosions have typically more slowly declining light curves.
For a given \nuc{56}{Ni} mass, however, the total ejecta mass will
also affect the light-curve width. We therefore propose
that the progenitors of fast declining SNe~Ia (most notably
1991bg-like objects) might be binaries of He and CO
WDs. The narrow light curves could then be understood as a natural
consequence of the smaller ejecta mass.  At the same time, the
spectral properties and colors of the explosion could be expected not
to change drastically compared to a merger of two CO WDs
with the same primary mass, since the secondary WD, which is burned
after the primary, only affects the central parts of the ejecta
\citep{pakmor2012a}.

\section{Brightness distributions, rates and delay times}
\label{sec:popsyn}

\begin{figure}
  \centering
  \includegraphics[width=\linewidth]{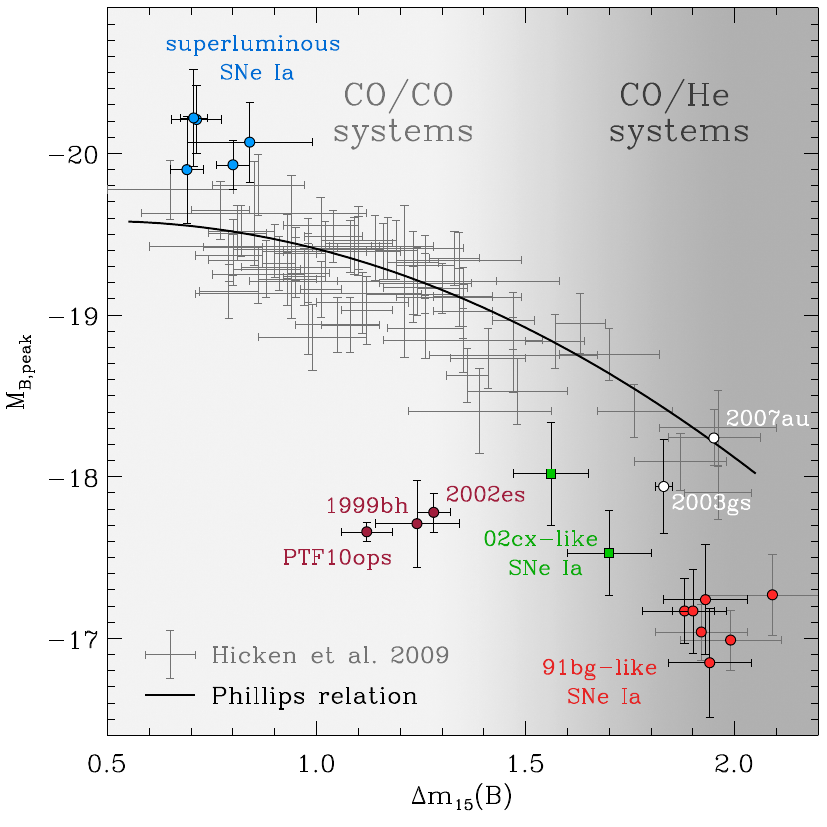}
  \caption{Observed diversity of SNe~Ia in \dm15 vs. 
    $M_{B,\mathrm{peak}}$ space as found in a sub-sample of the
    data presented in tables 8 and 9 of \citet{hicken2009b} where SNe 
    without a \dm15 measurement and a distance modulus $\mu \leq 
    33$\,mag have been removed. Most of these SNe~Ia are normal
    and follow the \citet{phillips1999a} relation. Additional datapoints
    \citep{phillips2007a,taubenberger2008a,taubenberger2011a,krisciunas2009a,
    maguire2011a,ganeshalingam2012a} have been included to indicate 
    the position of the peculiar sub-classes discussed in the text.
    Except for 2002cx-like SNe, our unified merger model potentially 
    explains the full observed diversity. Explosions in CO/CO binaries
    are expected to have more slowly declining light curves than 
    those in CO/He binaries (indicated by the background-color 
    gradient).}
  \label{fig:diversity}
\end{figure}

To explain the observed diversity of SNe~Ia with a theoretical model,
it is necessary to reproduce not only light curves and spectra of
single objects, but also the brightness distribution of objects of
different sub-classes, their relative rates, and their delay times.

In this regard, we base the analysis of our scenario on one specific
population-synthesis model that has been described in a set of papers
\citep{ruiter2009a,ruiter2011a,ruiter2012a}, since it is possibly the
most detailed model published at the moment. Note that other
population-synthesis models find somewhat different results,
which introduces significant uncertainties and may alter some of our
conclusions.

\citet{ruiter2012a} have shown that carbon-ignited violent mergers of
two CO WDs can not only explain the brightness
distribution of normal SNe~Ia but also their observed delay-time
distribution and rates. Moreover, they found a tendency that mergers
with more massive primary WDs, which lead to brighter supernovae,
preferentially occur in younger populations, consistent with observed
trends \citep[see e.g.][]{gupta2011a}. However, \citet{ruiter2012a}
find not enough systems to account for the observed number of faint
SNe. In nature those have predominantly fast light curves similar to 
SN~1991bg (see Fig.~\ref{fig:diversity}), but there seems to
emerge a sparse population with low luminosities yet slowly evolving
light curves similar to SN~2002es and PTF10ops
(\citealt{maguire2011a,ganeshalingam2012a}, see also
Fig.~\ref{fig:diversity}) which might be consistent with the mergers
of low-mass CO WD binaries.

In contrast to mergers of two CO WDs, the distribution of 
mass-transferring CO/He WD binaries has a strong peak between 
$0.8\,\msun$ and $0.9\,\msun$ for the mass of the primary WD 
\citep{ruiter2011a}. Although stable mass transfer is assumed for the
binary systems discussed in \citet{ruiter2011a}, we argue that a
significant fraction of them may cause a dynamical He detonation
on the surface of the primary WD \citep{guillochon2010a}, long before
a sufficient amount of He has been accumulated on the primary WD
to ignite due to thermal instabilities. Since a
dynamical detonation occurs, in this case, soon after mass transfer sets
in, the core mass of a CO WD in \citet{ruiter2011a} is still a good
indicator for the absolute brightness of the subsequent explosion. In
general, binaries with a He WD secondary should therefore lead to
fainter explosions since their primaries are typically significantly
less massive than those of CO WD binaries.
 
Mergers with a primary WD mass around $0.9\,\msun$ should produce a
\nuc{56}{Ni} mass close to that required for 1991bg-like SNe~Ia
\citep{pakmor2010a,sim2010a}.  For those primary masses, the rate of
potentially exploding CO/He WD binaries dominates
over that of pure CO WD binaries
\citep{ruiter2011a}. Since these systems will have narrow light
curves (see discussion above) they would naturally explain the
population of 1991bg-like SNe.  Moreover, the very long delay times of
these explosions (>\,1 Gyr; \citealt{ruiter2011a}) are in good
agreement with the observed tendency of 1991bg-like SNe to arise in
old stellar populations.

Population synthesis calculations also find a sparsely populated tail
of CO/He WD binaries towards larger primary masses
\citep{ruiter2011a}.  Explosions of such systems can reach
luminosities more typical for normal SNe~Ia but should still have
faster light curves similar to 1991bg-like SNe.
Observationally, these events might be identified with objects like
SN~2003gs or SN~2007au (\citealt{krisciunas2009a,hicken2009b}, 
see Fig.~\ref{fig:diversity}).

\begin{figure}
  \centering
  \includegraphics[width=\linewidth]{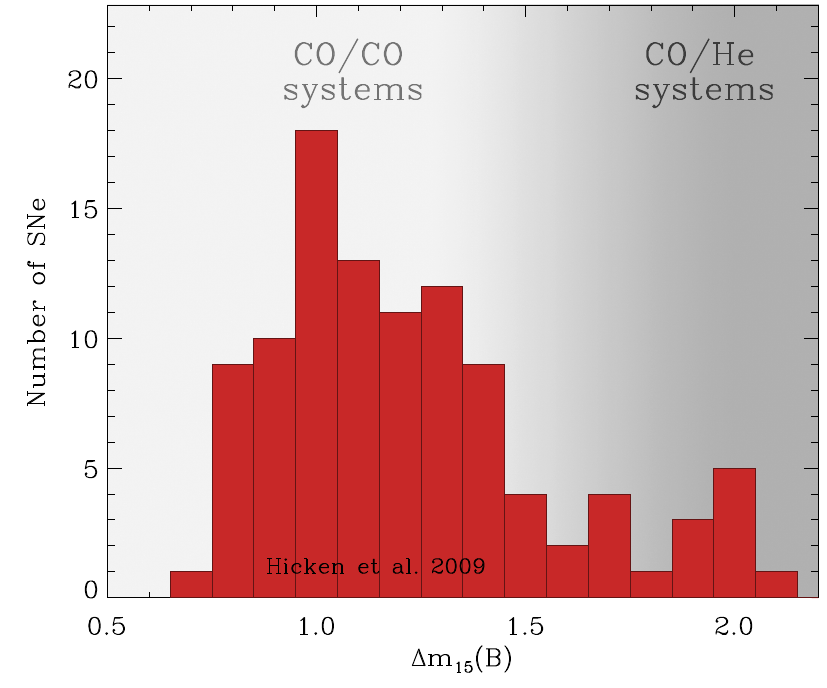}
  \caption{Histogram of the observed number of SNe~Ia as a function 
    of \dm15 (data from table 9 of \citet{hicken2009b}). The 
    distribution shows some indication for bimodality which might
    be identified with the two different companion populations 
    in our model (illustrated by the background-color gradient as 
    in Fig.~\ref{fig:diversity}).}
  \label{fig:dm15}
\end{figure}

It is interesting to note that the two populations might also provide a
natural explanation for the observed bimodal distribution of SNe~Ia in
\dm15-space which some SN data sets suggest (see Fig.~\ref{fig:dm15}).
The population synthesis calculations of \citet{ruiter2009a} predict a
total SN~Ia rate that is comparable to the Galactic SN~Ia rate if both
double-degenerate mergers with a total mass above the
Chandrasekhar-limit and single-degenerate Chandrasekhar-mass models
are considered. Hydrogen-accreting single-degenerate systems, however,
contribute only $\sim3$ per cent to this total rate. Therefore it
seems unlikely that the bimodal distribution in \dm15-space can be
attributed to contributions from single-degenerate and
double-degenerate progenitor systems.

\section{Summary and discussion}
\label{sec:summary}

In this paper, we have presented a new simulation of the merger of two
CO~WDs including thin He shells. We have shown that the presence
of only $0.01\,\msun$ of He is sufficient to ignite a detonation in
the He shell of the primary WD. This leads to a robust ignition
mechanism for CO~WD binaries that are about to merge. As in the
double-detonation explosion scenario for He-accreting
sub-Chandrasekhar-mass WDs, this He detonation subsequently sends
a converging shock into the underlying CO core which leads to a
detonation in the core and completely unbinds the
progenitor stars.

The new scenario works independently of the nature of the secondary
(He or CO) WD. We have argued that the two types of
progenitor systems lead to two distinct populations of SNe~Ia: mostly
faint, fast declining objects for He companions and bright, slowly
declining SNe for CO companions. This may naturally
explain the observed bi-modality of SNe~Ia in \dm15 space. We have
also shown that population synthesis qualitatively predicts brightness
distributions, relative rates and delay times for these two
populations in agreement with observational data of normal
and subluminous SNe~Ia.

Other observational sub-classes can be easily incorporated
into our scenario. SNe~Ia with signs of CSM interaction \citep[see, 
e.g.,][]{patat2007a,dilday2012a} can be interpreted as double-degenerate 
systems exploding shortly after the last common-envelope phase 
\citep{livio2003a,soker2013a}. Superluminous SNe~Ia, 
which are significantly brighter than normal SNe~Ia, have recently been 
argued to originate from a merger of two massive WDs
without a prompt detonation \citep{hachinger2012a,taubenberger2013a}. 
In our scenario these might be explained by very He-poor binary systems 
with two massive CO~WDs.

The only objects that can probably not be accounted for by pure 
detonations are 2002cx-like SNe \citep{jha2006a}. But they have 
recently been identified with off-center-ignited pure deflagrations 
of Chandrasekhar-mass CO WDs \citep{kromer2013a}.

Together, our scenario may provide a first step to construct a comprehensive
picture of SNe~Ia that is in principle able to explain all major sub-classes
including superluminous SNe~Ia and SNe~Ia with CSM interaction.

\begin{acknowledgements}
  We thank A. Ruiter, M. Fink, S. Hachinger, W. Hillebrandt, 
  F. R\"opke, I. Seitenzahl and S. Sim for many inspiring discussions. 
  R.P. and M.K. are grateful for IPMU's hospitality during a research
  visit at which this work was initiated.
  This work was supported by the Klaus Tschira Foundation, the DFG
  through the cluster of excellence `Origin and Structure of the
  Universe' (EXC~153) and the Transregional Collaborative Research
  Centre `The Dark Universe' (TRR~33) and in part by WPI Initiative, 
  MEXT, Japan.
\end{acknowledgements}

\bibliographystyle{apj}

\end{document}